\documentclass[conference]{IEEEtran}
\IEEEoverridecommandlockouts

\usepackage{tikz}
\nonstopmode
\usepackage{cite}
\usepackage{amsmath,amssymb,amsfonts}
\usepackage{graphicx}
\usepackage{textcomp}
\usepackage{xcolor}

\usepackage{commath}
\usepackage{algorithm}
\usepackage{algpseudocode}
\usepackage{optidef}
\usepackage[symbol]{footmisc}
\usepackage{enumitem}
\usepackage{svg}
\usepackage{lettrine}
\usepackage{subcaption}
\usepackage{float}
\usepackage{caption}

\usepackage{geometry}
\IEEEaftertitletext{\vspace{-0.5\baselineskip}}

\makeatletter
\newcommand{\nosemic}{\renewcommand{\@endalgocfline}{\relax}}
\newcommand{\dosemic}{\renewcommand{\@endalgocfline}{\algocf@endline}}

\let\oldnl\nl
\newcommand{\nonl}{\renewcommand{\nl}{\let\nl\oldnl}}
\makeatother

\renewcommand{\arraystretch}{1.25}

\def\BibTeX{{\rm B\kern-.05em{\sc i\kern-.025em b}\kern-.08em
    T\kern-.1667em\lower.7ex\hbox{E}\kern-.125emX}}

\begin{document}

\title{Vision Transformer Based Semantic Communications for Next Generation Wireless Networks
}
\author{
    \IEEEauthorblockN{
        Muhammad Ahmed Mohsin\IEEEauthorrefmark{1}, Muhammad Jazib\IEEEauthorrefmark{2}, Zeeshan Alam\IEEEauthorrefmark{3}, 
        Muhmmad Farhan Khan\IEEEauthorrefmark{4},
        Muhammad Saad\IEEEauthorrefmark{5},\\ Muhammad Ali Jamshed\IEEEauthorrefmark{6}
       }

        \IEEEauthorblockA{\IEEEauthorrefmark{1}Department of Electrical Engineering, Stanford University, Stanford, CA, USA}

    \IEEEauthorblockA{\IEEEauthorrefmark{2}Department of Electrical Engineering (DEE), PIEAS, Pakistan}
     \IEEEauthorblockA{\IEEEauthorrefmark{3}Faculty of Computer Science,  University of New Brunswick,
NB, Canada}

\IEEEauthorblockA{\IEEEauthorrefmark{4}School of Computer Science and Information Technology, University College Cork, Cork, Ireland}

\IEEEauthorblockA{\IEEEauthorrefmark{5}School of Electrical Engineering and Computer Science (SEECS), NUST, Pakistan}

\IEEEauthorblockA{\IEEEauthorrefmark{6}College of Science and Engineering, University of Glasgow, UK}
    
    \IEEEauthorblockA{Email: \{muahmed@stanford.edu, bsee2023@pieas.edu.pk, Muhammad.alam@unb.ca, \\Farhan.khan@cs.ucc.ie, msaadbee20seecs@seecs.edu.pk, muhammadali.jamshed@glasgow.ac.uk \}
    }



}

\maketitle

\begin{abstract}
In the evolving landscape of 6G networks, semantic communications are poised to revolutionize data transmission by prioritizing the transmission of semantic meaning over raw data accuracy. This paper presents a Vision Transformer (ViT)-based semantic communication framework that has been deliberately designed to achieve high semantic similarity during image transmission while simultaneously minimizing the demand for bandwidth. By equipping ViT as the encoder-decoder framework, the proposed architecture can proficiently encode images into a high semantic content at the transmitter and precisely reconstruct the images, considering real-world fading and noise consideration at the receiver. Building on the attention mechanisms inherent to ViTs, our model outperforms Convolution Neural Network (CNNs) and Generative Adversarial Networks (GANs) tailored for generating such images. The architecture based on the proposed ViT network achieves the Peak Signal-to-noise Ratio (PSNR) of 38 dB, which is higher than other Deep Learning (DL) approaches in maintaining semantic similarity across different communication environments. These findings establish our ViT-based approach as a significant breakthrough in semantic communications.

\end{abstract}


\section{Introduction}
\lettrine{S}{emantic} communications, a new paradigm in wireless communication, intends to transmit the essence of the information instead of merely the data. This method is especially crucial in 6G networks, where the importance of achieving high bandwidth efficiency and the ability to sustain the connection in adverse conditions is significant~\cite{jamshed2024non}. Conventional methods of sending information , which largely depends on bit-level transmissions are often unable to meet specific standards for optimal bandwidth utilization or signal stability in the presence of interference. 

The advent of Deep Learning (DL) architectures has significantly advanced the fields of computer vision, image processing, and wireless communication. Several studies have proposed methods for efficient bandwidth utilization in this context\cite{10847914}. Autoencoders have long been used for compressed communication. A robust technique proposed in~\cite{8820761} combined a basic autoencoder with a denoising autoencoder, yielding remarkable results. However, this study only investigates the impacts of quantization and Additive White Gaussian Noise (AWGN). Similarly,~\cite{zebang2019densely} proposed a densely connected autoencoder structure, inspired by the DenseNet architecture, to maximize feature extraction. They also developed a U-Net-like structure to reduce distortion. Although their method outperformed JPEG 2000, some artifacts were still observed.

The evolution from 5G to 6G requires better approaches that consume less bandwidth than the current typical ones and maintain the transmitted semantic content. In particular, the capabilities of DL to extract and prioritize useful features, makes it highly suitable for semantic communication. Specifically, in computer vision, solutions like Convolutional Neural Networks (CNNs) show promise for compressing and transmitting image data~\cite{xie2021deep}. 

CNNs have traditionally excelled at image compression and transmission due to their strengths in local feature extraction and spatial hierarchies. In~\cite{saber2022list}, authors explored CNN-based applications for image transmission in noisy channels with channel coding and denoising. On one hand, CNNs are limited by their local receptive fields, which impairs their ability to capture global context important aspect of semantic communication. On the other hand, Vision Transformers (ViTs) have recently emerged as an effective alternative to CNNs, with greater semantic communication capabilities. ViTs can encode global context and transmit more semantic features, making them well-suited for bandwidth-limited and channel-interference environments. Unlike more conventional methods, ViTs are focused on maintaining semantic communication.


This research explores the potential of ViTs in semantic communication by contrasting their performance with that of CNNs and Generative Adversarial Networks (GANs) on datasets including ImageNet, CIFAR-10, and CIFAR-100. We assess the resilience of these models under challenging channel conditions, such as AWGN, Rayleigh fading, Rician fading, and Nakagami-m fading.

Inspired by the advantages of ViTs, our paper makes the following contributions to knowledge:
\begin{itemize}
    \item Presented a custom ViT architecture for efficient semantic transmission, surpassing the efficacy of state-of-the-art autoencoders. 
    \item Conducted a performance analysis across multiple datasets, simulating various fading models to accurately replicate real-world wireless environments.
    \item Compared the accuracy of ViT against multiple DL models to evaluate robustness in relation to contemporary semantic models, achieving 72\% less bandwidth utilization.
\end{itemize}

\section{System Model}
\begin{figure}[t!]
    \centering
    \includegraphics[width=\linewidth]{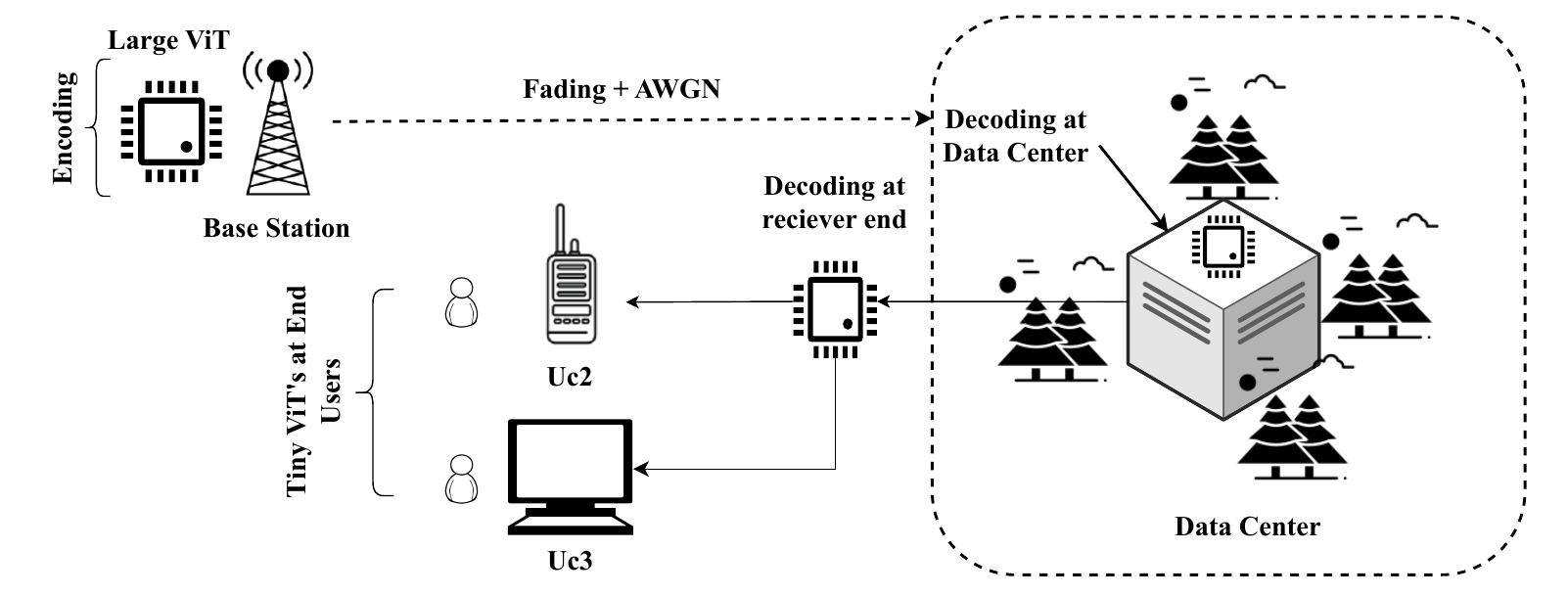}
    \caption{System model of proposed wireless network system}
    \label{fig:sysmodel}
\end{figure}

As shown in the Fig.~\ref{fig:sysmodel}, we assume a system model for compressed communication, particularly semantic communication, using ViT architecture consisting of large and small ViT models. We assume processing of input images \(X_u\) at the transmitter side. These images are sliced into \(N\) patches, with each patch confined to linear embedding and converted to suitable vector representations. These embedded patches are then fed into a ViT-based encoder, where they are encoded into learnable features \(\mathbf{F}\).
\subsection{BPSK Modulation}
Binary Phase Shift Keying (BPSK) modulation is a scheme used to transmit  the message by modulating the phases of the reference signal \cite{dayana2021analysis}. In our model, binary data streams \(x_i(t)\) from the ViT model were processed through a bipolar NRZ encoder, mapping \(0\) and \(1\) to voltage levels \(-1\) and \(+1\), forming \(X(t)\). This waveform was modulated using BPSK, shifting the carrier phase by \(0^\circ\) for \(+1\) and \(180^\circ\) for \(-1\).

As BPSK signal is a phase shift keying (PSK) signal where the phase of the signal varies depending on the input bit at a particular instant:

\begin{equation}
\begin{aligned}
X_{\text{PSK}}(t) &= A \cos(\omega_c t) \quad \text{for input bit 1}, \\
\end{aligned}
\end{equation}

\begin{equation}
\begin{aligned}
X_{\text{PSK}}(t) &= A \cos(\omega_c t + \pi) \quad \text{for input bit 0}.
\end{aligned}
\end{equation}
 
With only two phases, it can allow the transmission of data while using as little bandwidth as possible, which is essential and in line with the processing demands of ViTs. Here, each bit \(b_i\) is converted into a symbol using the formula:
\begin{equation}
s_i = 2b_i - 1,
\end{equation}
\noindent where \(b_i\) is either 0 or 1. This modulation process transforms the binary features into symbols ready for transmission.
\subsection{Communication Channel}
\begin{figure}[t!]
    \centering
    \includegraphics[width=\linewidth]{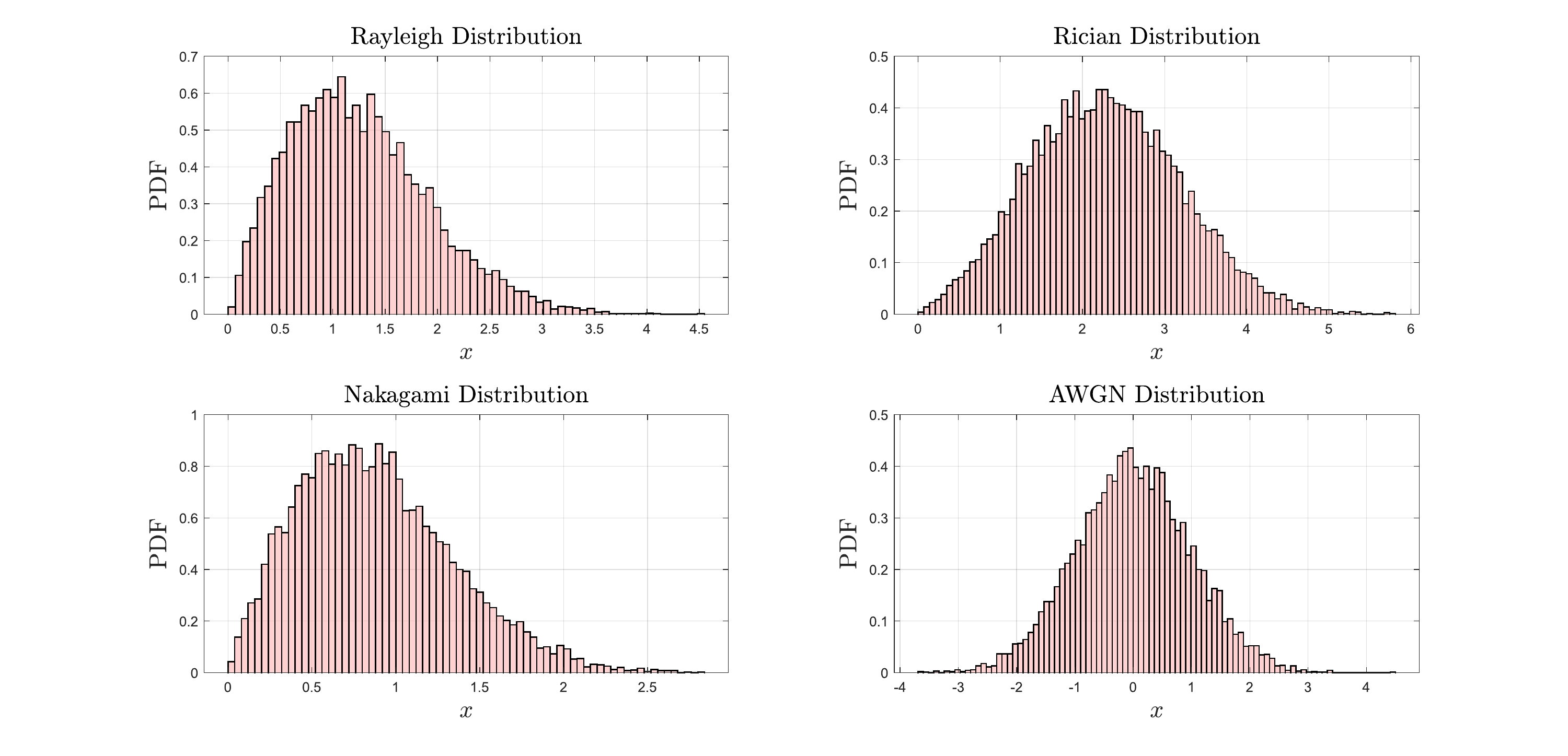}
    \caption{Sampled probability density function (PDF's) of fadings}
    \label{fig:pdfs}
\end{figure}

To mimic the stochastic nature of real-world communication scenarios, we incorporated both Line-of-Sight (LOS) and Non-Line-of-Sight (NLOS) channel characteristics~\cite{10464446}. Firstly, their Probability Density Functions (PDFs) were plotted as shown in Fig.~\ref{fig:pdfs}, and then those distributions were sampled. For Rayleigh and Rician fading channels, we generated complex Gaussian random variables \( h_I \) and \( h_Q \) with zero mean and unit variance. The Rayleigh channel can be expressed as:
\begin{equation}
h(t) = \frac{1}{\sqrt{2}} \left( h_I(t) + jh_Q(t) \right),
\end{equation}

\noindent where the  PDF of the Rayleigh distributed amplitude is given by \( f_R(r) = \frac{r}{\sigma^2} e^{-\frac{r^2}{2\sigma^2}} \), and the channel modeling is further enhanced by including the LOS component. The Rician channel model with fading coefficient \( h(t) \) is given by:
\begin{equation}
h(t) = \sqrt{\frac{K}{K+1}} + \sqrt{\frac{1}{K+1}} \cdot \frac{1}{\sqrt{2}} \left( h_I(t) + jh_Q(t) \right),
\end{equation}

where the PDF of Rician distribution amplitude is given by \( R(r) = \frac{r}{\sigma^2} e^{-\frac{r^2 + \nu^2}{2\sigma^2}} I_0\left(\frac{r\nu}{\sigma^2}\right) \), where \( \nu = \frac{2K}{\sigma^2} \) and \( I_0(\cdot) \) is the bessel function.

Nakagami-\( m \) being more general representation of fading scenarios was also incorporated to mimic all the statistical channel fading scenarios. The PDF of the Nakagami-\( m \) distributed amplitude \( r \) is given by \( R(r) = \frac{2m^m}{\Gamma(m)\Omega^m} r^{2m-1} e^{-\frac{m r^2}{\Omega}} \), with \( \Gamma(m) \) is the Gamma function, \( m \geq \frac{1}{2} \) is the shape parameter, and \( \Omega = \mathbb{E}[r^2] \) is the spread parameter, the mean power of the fading envelope. 

The presence of thermal noise in electronic communication systems is very common. To emulate this AWGN, stationary noise was added. The PDF of Gaussian noise \( n \) is given by \( N(n) = \frac{1}{\sqrt{2\pi\sigma^2}} e^{-\frac{n^2}{2\sigma^2}} \).

The modulated signals \(s_i\), were then transmitted over a communication channel that was carried with small scale fading to mimic the real communication environment. The received signal \(y\) can be modeled as:
\begin{equation}
y = h \cdot s + n,
\end{equation}
\noindent where \(h\) represents the channel fading coefficients, \(s\) is the transmitted BPSK signal, and \(n\) is the AWGN. Inclusion of fading and noise in the system model leverages both LOS and NLOS scenarios to study the effects more effectively.

\subsection{Feature to Bit Conversion}
Data center possibly receiving encoded data from multiple sources was taken into consideration. The received signal \(y\) at the data center is first demodulated to retrieve the encoded features, which are then processed by a decoder to reconstruct the original data. IEEE 754 representation has been used to represent the floating-point numbers in the binary format. Single precision (32 bits) has been taken into account for this conversion~\cite{7203811}. Statistical methods were incorporated for element-wise conversions. The vector size decorator was configured for this efficient conversion from integer to floating point and floating to bit conversions, and vice versa, by compiling the decorated functions to machine code.

The IEEE 754 format comprises three components: the sign bit \( S \), the exponential value \( E \), and the mantissa \( M \). The value \( V \) of the floating point is given by:
\begin{equation}
V = (-1)^S \times M \times 2^E.
\end{equation}

\subsection{Data Sets}
To have a comprehensive evaluation of our DL models, we experimented with three different datasets to train our proposed  DL Models. At the beginning, we trained the DL models using Cifar-10, consisting of 32$\times$32 colored images with labels against these classes. To further assess the robustness of our model, we also integrated another dataset, Cifar-100, which consists of 100 classes. Finally, we evaluated the scalability and performance of our proposed custom model against a documented baseline on the imagenette dataset that contains more than one million images from a thousand classes.
\section{Deep Learning Models}
\subsection{Transformer Architecture}
We propose a custom ViT model, composed of following parts: (i) \textit{Non-Sequential:} Unlike traditional CNNs, which process images sequentially, ViTs handle images differently. Their non-sequential nature allows them to acquire global context and enables more parallelization, resulting in reduced training time compared to CNNs. (ii) \textit {Self Attention:} Ranking the similarity scores between different patches in an image, processing through sequential data. (iii) \textit {Positional Embedding:} As ViTs are non-sequential learning models, positional embedding are utilized to restore the spatial arrangement of image patches. Our Denoising Autoencoder Vision Transformer model (DAE-ViT) consists of three modules primarily: (a) encoder, (b) decoder, and (c) additional utility (fading channels). We experimented ViT model by formulating ViT base model, its scales are visible as shown in the Table ~\ref{tab:dae_vit_config}.

\begin{table}[b!]
    \centering
    \caption{Configuration of DAE-ViT Models}
    \label{tab:dae_vit_config}
    \resizebox{0.99\columnwidth}{!}{%
        \begin{tabular}{|c|c|c|c|}
            \hline
            \textbf{Configuration} & \textbf{ViT Small} & \textbf{ViT Base} & \textbf{ViT Large} \\
            \hline
            \hline
            Patch Size         & 16  & 16  & 16  \\
            Embedding Dimension & 384 & 768 & 1024 \\
            Encoder Layers     & 12  & 12  & 24  \\
            Encoder Heads      & 6   & 12  & 16  \\
            Decoder Layers     & 8   & 8   & 16  \\
            Decoder Heads      & 8   & 16  & 16  \\
            \hline
        \end{tabular}%
    }
\end{table}
\begin{figure*}[h!]
    \centering
    \includegraphics[width=\textwidth]{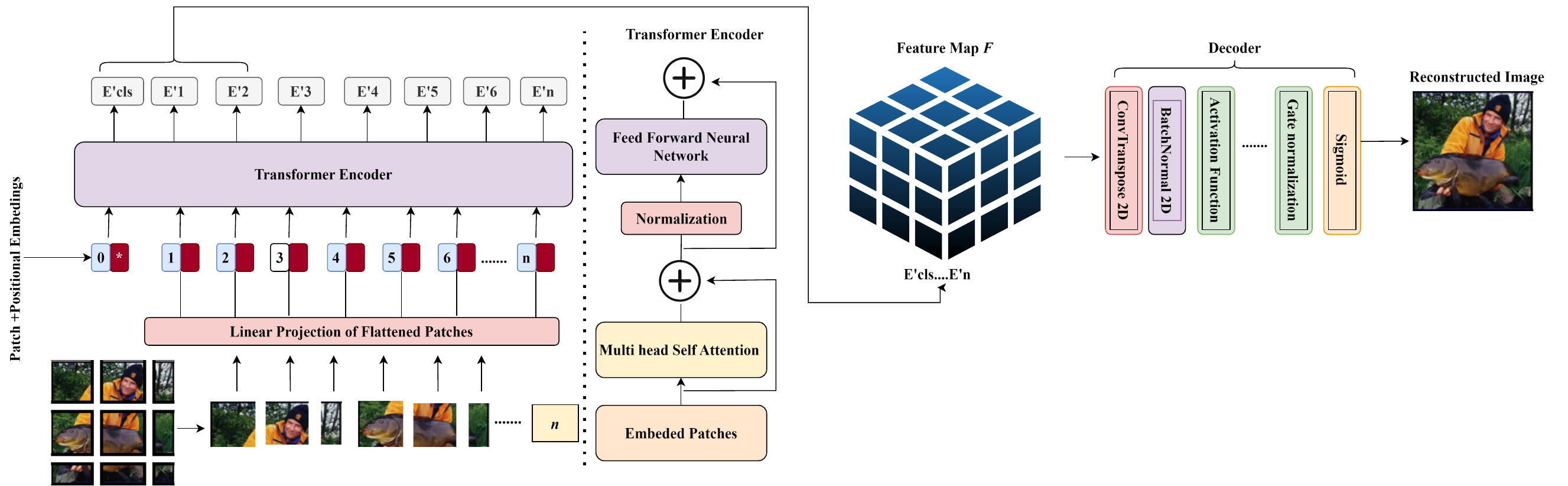}
    \caption{Framework of proposed ViT architecture }
    \label{fig:net}
\end{figure*}
However, different models can also be implemented in different ways by adjusting the hyperparameters to adapt the model complexity of different conditions by following the algorithm~\ref{alg:pd_noma}.

\begin{algorithm}[t!]
    \caption{Algorithm for Semantic Communication}\label{alg:pd_noma}
    \begin{algorithmic}[1]
        \State \textbf{Input:} \parbox[t]{\dimexpr\linewidth-\algorithmicindent}{Image \( X \) for encoding with \( x \times y \) dimensions.}
        
        \vspace{0.1em}  
        \State \textbf{Step 1:} \parbox[t]{\dimexpr\linewidth-\algorithmicindent}{Image Compression.}
        
        \For{\( \text{epoch} = 1 \) \textbf{to} \( N \)}
            \State \parbox[t]{\dimexpr\linewidth-\algorithmicindent}{Obtain positional embedding of \( X \) with dimension \( \theta_e \).}
            \State \parbox[t]{\dimexpr\linewidth-\algorithmicindent}{Obtain patches of \( X \) with dimensions \( \theta_p \).}
            \State \parbox[t]{\dimexpr\linewidth-\algorithmicindent}{Pass the input image \( X \) through transformer to extract global features \( \theta_f \) with dimension \( (\theta_e, \theta_n, \theta_l) \).}
            \vspace{0.1em}
        \EndFor

        \State \textbf{Step 2:} Phase Modulation and Interference Management.
        
        \State Perform BPSK modulation to convert features to symbols.
        
        \State Add pathloss \( \theta_z \) and fading \( \theta_w \) to the encoded features \( \theta_f \) to mimic the wireless environment.

        \State \textbf{Step 3:} Image Reconstruction.
        
        \For{\( \text{epoch} = 1 \) \textbf{to} \( N \)}
            \State \parbox[t]{\dimexpr\linewidth-\algorithmicindent}{Pass the noisy features through decoder positional embedding \( \theta_e \) to obtain the position of patches \( \theta_x \).}
            \vspace{0.01em}
            
            \State \parbox[t]{\dimexpr\linewidth-\algorithmicindent}{Pass the output \( \theta_x \) through the transformer block to remove the noise and reconstruct the image.}
            \vspace{0.01em}
            
            \State \parbox[t]{\dimexpr\linewidth-\algorithmicindent}{Arrange the patches in order to get the denoised reconstructed image from the features \( \theta_f \).}
        \EndFor

        \State \textbf{Output:} Reconstructed denoised image \( \mathbf{X} \in \mathbb{R}^{H \times W \times C} \).
    \end{algorithmic}
\end{algorithm}

\textbf{Encoder:}
The input that is passed into ViT is patch units of images. These patches are transformed into a two-dimensional sequence, called embedded patches $X \in \mathbb{R}^{(N + 1) \times D}$, to comprehend the reciprocity between them utilizing Multi-Head Self Attention (MHSA) operations. MHSA is basically a part of transformer model that makes the model capable of focusing on different parts of input sequences simultaneously~\cite{vaswani2017attention}.

Input images are initially passed through the patch layer and divided into patches of $16\times16$ processing, pixels, where the dimension of each image is set to be $224\times224$ pixels. Each patch is then embedded into a vector with a specified dimension of 768.

The input images $X_u$ are divided into small non-overlapping patches of size $(S \times S)$, where the number of patches is $N = \frac{w}{S^2}$, and $w = l \times N_c$. The patches are transformed into vectors $x_{p_{u,j}} \in \mathbb{R}^{S^2}$, for $(1 \leq j \leq N)$ to be incorporated into the model dimension $d$, where a linear projection $E \in \mathbb{R}^{S^2 \times d}$ is utilized. The output of the patch is referred to as patch embedding. A class token $x_{cls}$, is inserted into the embedded patches. 

Afterward, positional embedding $E_{pos} \in \mathbb{R}^{(N + 1) \times d}$, are added to encipher the sequence of input. They are initialized as a learnable parameter with the same number of positions as the number of patches~\cite{alexey2020image}. The output of the patch and position embedding $Z_0$ is given by:
\begin{equation}
Z_0 = [x_{cls}; x_{p_{u,1}}E; x_{p_{u,2}}E; \ldots; x_{p_{u,N}}E] + E_{pos}.
\end{equation}

Attention weight matrix (A) calculates attention scores between tokens to direct the information aggregation process. The compatibility function was applied to calculate the attention score\textbf{$A_{ij}$} between query $q_i$ and key $k_j$~\cite{li2022variable}. The weighted sum of all elements value $v$ of patch embedding $X$ reflects the attention information in the current embedding. The concatenated query, key, and value vectors  \([q, k, v]\)is given by:
\begin{equation}
[q, k, v] = XU, \quad U \in \mathbb{R}^{D \times 3D}_h.
\end{equation}

The learned weight matrix \( U_{qkv} \) associates the vectors \( x \) with the query \( q \), key \( k \), and value \( v \) containing dimensions  \( D \times 3D_h \), where \( D \) denotes the dimensionality of the input embedding and \( 3D_h \) stands for the concatenated dimensions of \( q \), \( k \), and \( v \).

The attention weights computation~\cite{vaswani2017attention} directs the flow and processing of information in the model and was eventually calculated through:

\begin{equation}
A = \text{softmax}\left( \frac{qk^{\top}}{\sqrt{D_h}} \right), \quad A \in \mathbb{R}^{(N+1) \times (N+1)},
\end{equation}

\noindent where $\text{softmax}(\mathbf{z})_i$ is defined as $\text{softmax}(\mathbf{z})_i = \frac{e^{z_i}}{\sum_{j} e^{z_j}}$.

The integrated embedding, patch embedding, positional embedding, and class token were rearranged in a shape to further process it through a series of transformer blocks, where each transformer block is composed of MHSA and feed-forward layers, making the model learn to capture local and global dependencies across the image. The equation for MHSA is given by:
\begin{equation}
\begin{gathered}
\text{MHSA}(X) = [SA_1(X); SA_2(X); \ldots; SA_k(X)] U_{\text{msa}}, \\
U_{\text{msa}} \in \mathbb{R}^{k \cdot D_h \times D}.
\end{gathered}
\end{equation}

MHSA(X) applied to $X$ where each \( SA_i(X) \) computes self-attention independently. \( U_{msa} \) is a learned weight matrix that combines the outputs of all attention heads into a final output, with dimensions \( k \cdot D_h \times D \).

The output from each transformer block was then processed through layer normalization before taking output. This normalization step makes the process of stabilization and improvements in the training procedure easier, since it ensures that the permutations of the variance and the mean are identical across the embedding dimensions.

\textbf{Decoder:}
The ViT acquires the features $F$ generated by the encoder and reconstructs them into an image \ensuremath{F \in \mathbb{R}^{N^{1/2} \times N^{1/2} \times D}}. Transformer blocks in the decoder were leveraged to process those features, which generated the images later~\cite{wu2024generalized}. Subsequently, positional embedding of vector size 768 was added for 196 different positions of image patches and one for the class token. This step was done to retain the spatial information of the patches in the image.

Furthermore, the combined embedding was then proceeded through transformer blocks, where the local and global dependencies in the data were captured using the MHSA mechanism. The original input was summed up with this output. Each position was then applied with a feed-forward neural network, and the output of the neural network was then normalized. The features were then rearranged into sequences, and the class token was then removed. Each feature vector is employed with a linear layer to be converted into a patch-sized vector of 768 pixels. These patches were rearranged to an image to get the full image: \ensuremath{\hat{X} \in \mathbb{R}^{H \times W \times C}}, as shown in Fig.~\ref{fig:output}. Lastly, the sigmoid gate function was implemented in order to retain high fidelity while minimizing artifacts and noise.

\subsection{Convolutional Neural Networks(CNN)}
Particularly, traditional CNN layers with 2D CNN modules were incorporated to transform learned features \( s \) in \( X \in \mathbb{R}^{B \times F \times 2N} \), where \( B \) is the batch size and \( F \) is the number of frames. The proposed model comprised of Principal Component Analysis (PCA) for dimensionality reduction to decompose the feature matrices \( W \) and \( \hat{W} \) to minimize the mean square error:

\begin{equation}
\| (X W) \hat{W} - X \|^2_2 \rightarrow_{W, \hat{W}} \min.
\end{equation}

For the original \( n \) samples, direct \( W \in \mathbb{R}^{m \times d} \) and reverse \( \hat{W} \in \mathbb{R}^{d \times m} \) transformations were employed, where a dense layer was a fully connected layer with linear activation:

\begin{equation}
f(X) = W \cdot X + \vec{b}.
\end{equation}

For encoding, convolutional and pooling layers were stacked with dense layers. The output was flattened before being added to the dense layer~\cite{simonyan2014very}. For semantic decoding, transpose convolution was adopted to aggregate patches of an image. Input data was passed with channel fading and AWGN noises.

\begin{figure}[t!]
     \centering
     \begin{subfigure}[t]{0.4\columnwidth} 
         \centering
         \includegraphics[width=\textwidth]{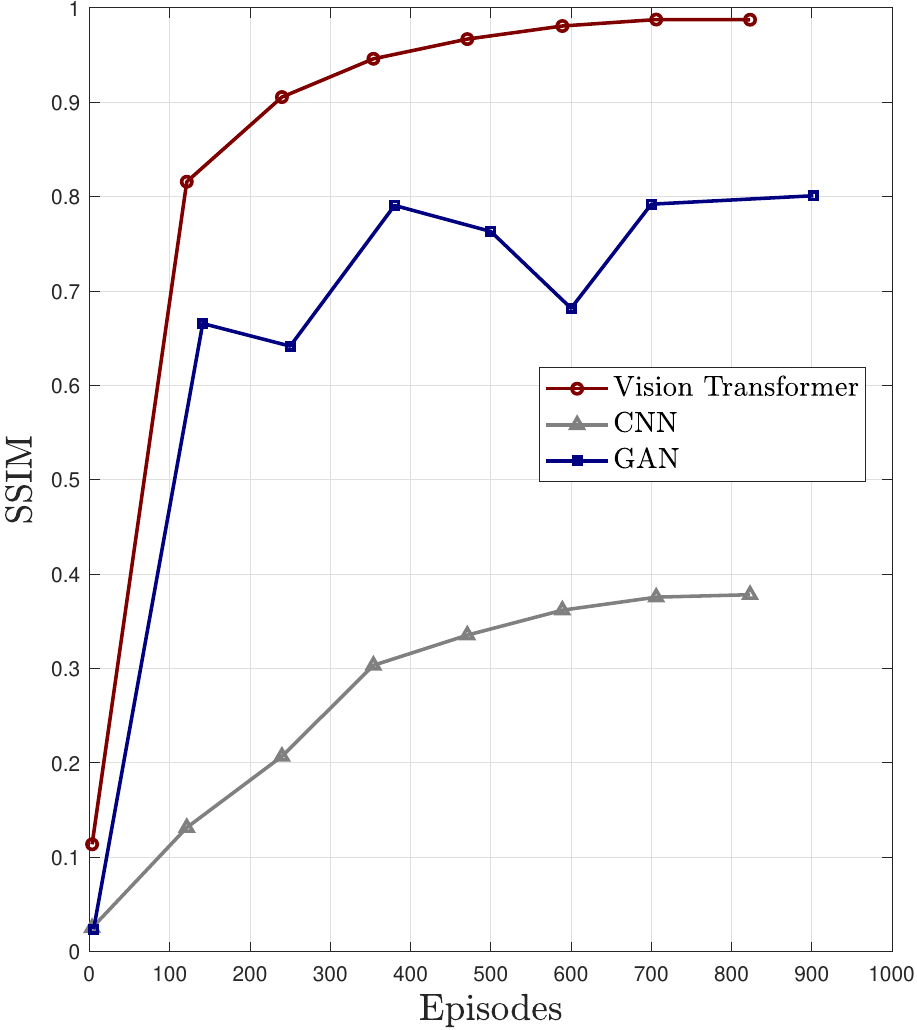}
         \caption{PSNR}
         \label{PSNR}
     \end{subfigure}
     \hspace{0.02\columnwidth} 
     \begin{subfigure}[t]{0.39\columnwidth} 
         \centering
         \includegraphics[width=\textwidth]{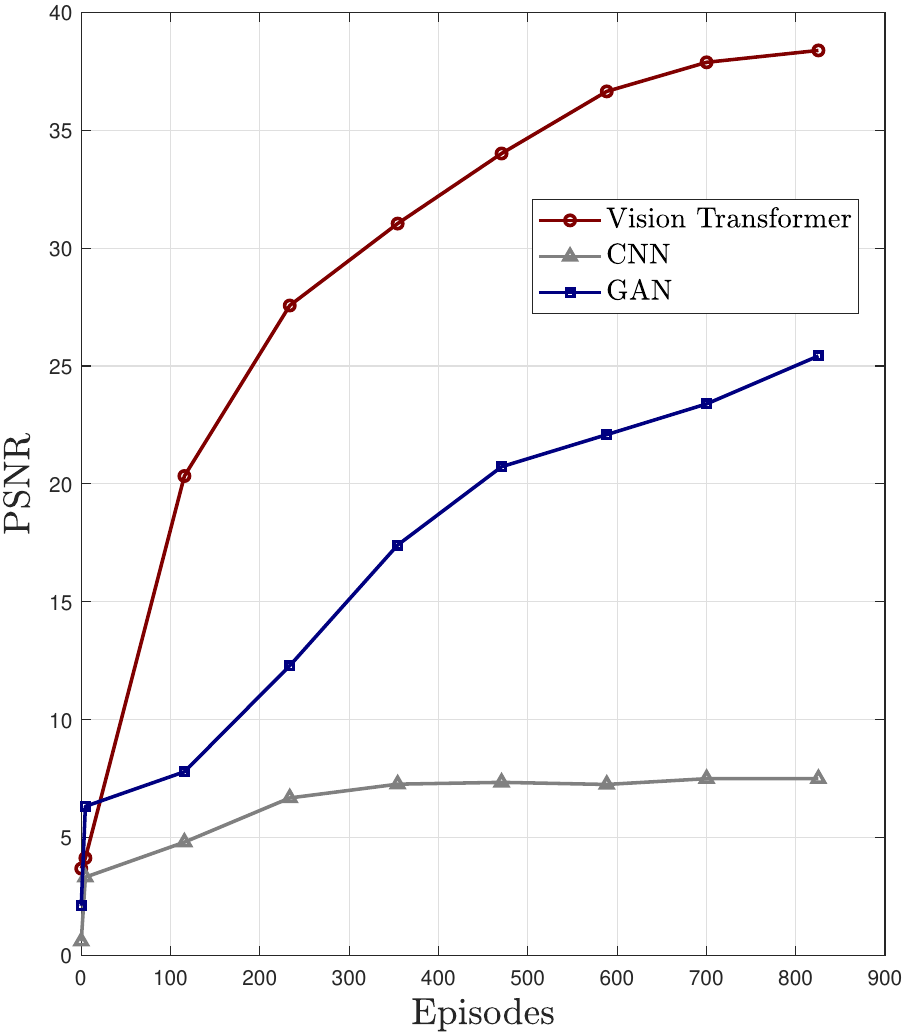}
         \caption{SSIM}
         \label{SSIM}
     \end{subfigure}
     \caption{Image quality under different fading scenarios}
     \label{fig:Image quality under different fading scenarios}
\end{figure}

\subsection{Generative Adversarial Networks(GANs)}
GANs were employed due to their ability to regenerate rich images~\cite{lokumarambage2023wireless}. The generator, \( G \), maps samples \( \hat{w} \) from a fixed known distribution \( p_{\hat{w}} \) to an unknown joint distribution \( p_{x|s} \), facilitating semantic reconstruction. The discriminator \( D \) differentiates between real inputs \((x, s)\) and generated inputs \((G(\hat{w}, s), s)\). The architecture is governed by the following conditional functions:
\begin{equation}
L_G = \mathbb{E}_{\hat{w} \sim p_{\hat{w}}} \left[ -\log(D(G(\hat{w}, s), s)) \right],
\end{equation}

\begin{equation}
L_D = - \mathbb{E}_{\hat{w} \sim p_{\hat{w}}} \log(1 - D(G(\hat{w}, s), s)) 
- \mathbb{E}_{x \sim p_{x|s}} \log D(x, s)
\end{equation}

The adversarial loss \( L_G \) represents the distribution loss inherent to GANs. Additionally, the distortion can be quantified as the L1 loss of the images:
\begin{equation}
\mathbb{E}[\|G(\hat{w}, s) - x\|_1].
\end{equation}

\begin{figure}[t!]
     \centering
     \begin{subfigure}[t]{0.4\columnwidth} 
         \centering
         \includegraphics[width=\textwidth]{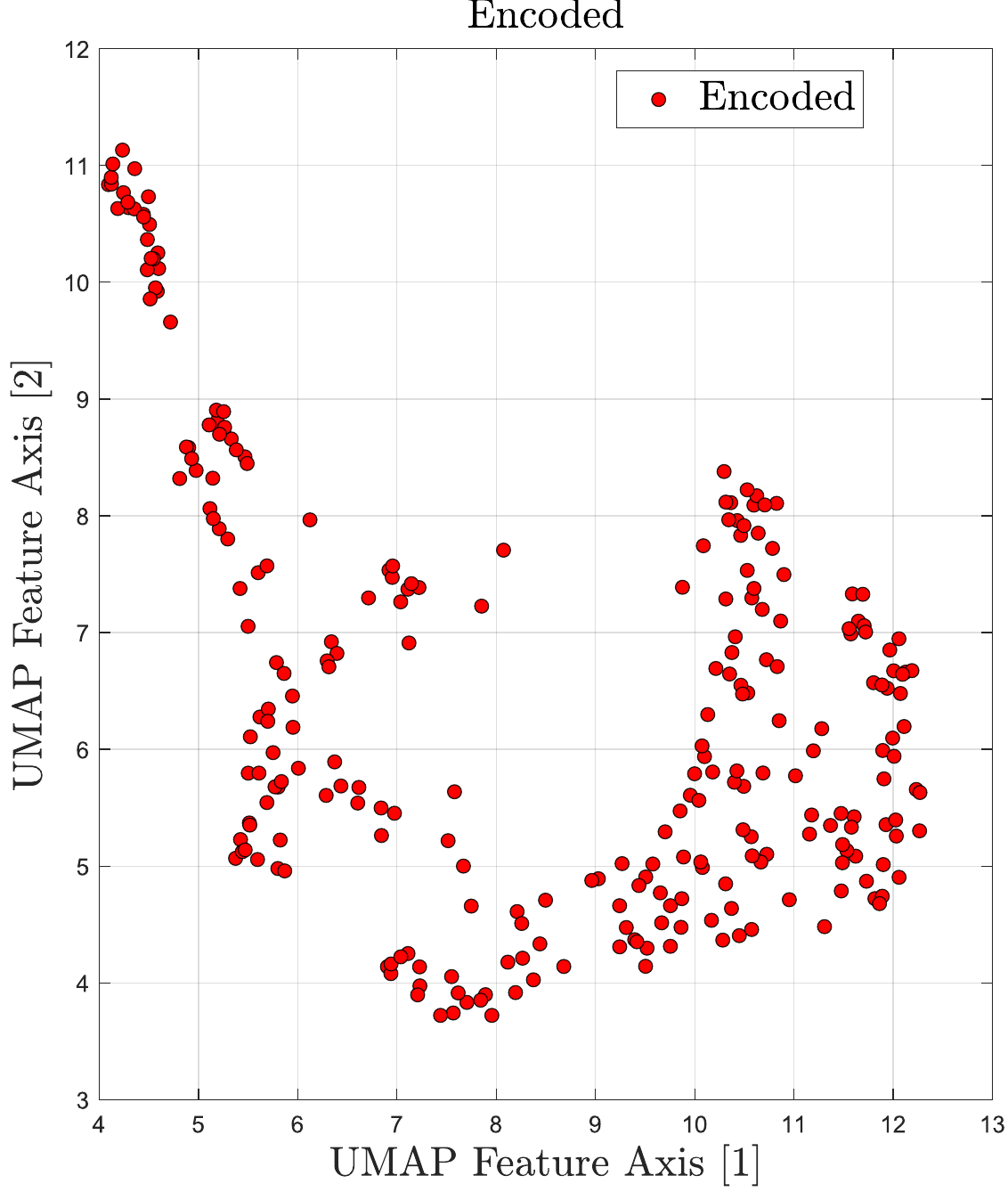}
         \caption{Encoded}
         \label{Encoded}
     \end{subfigure}
     \hspace{0.02\columnwidth} 
     \begin{subfigure}[t]{0.4\columnwidth} 
         \centering
         \includegraphics[width=\textwidth]{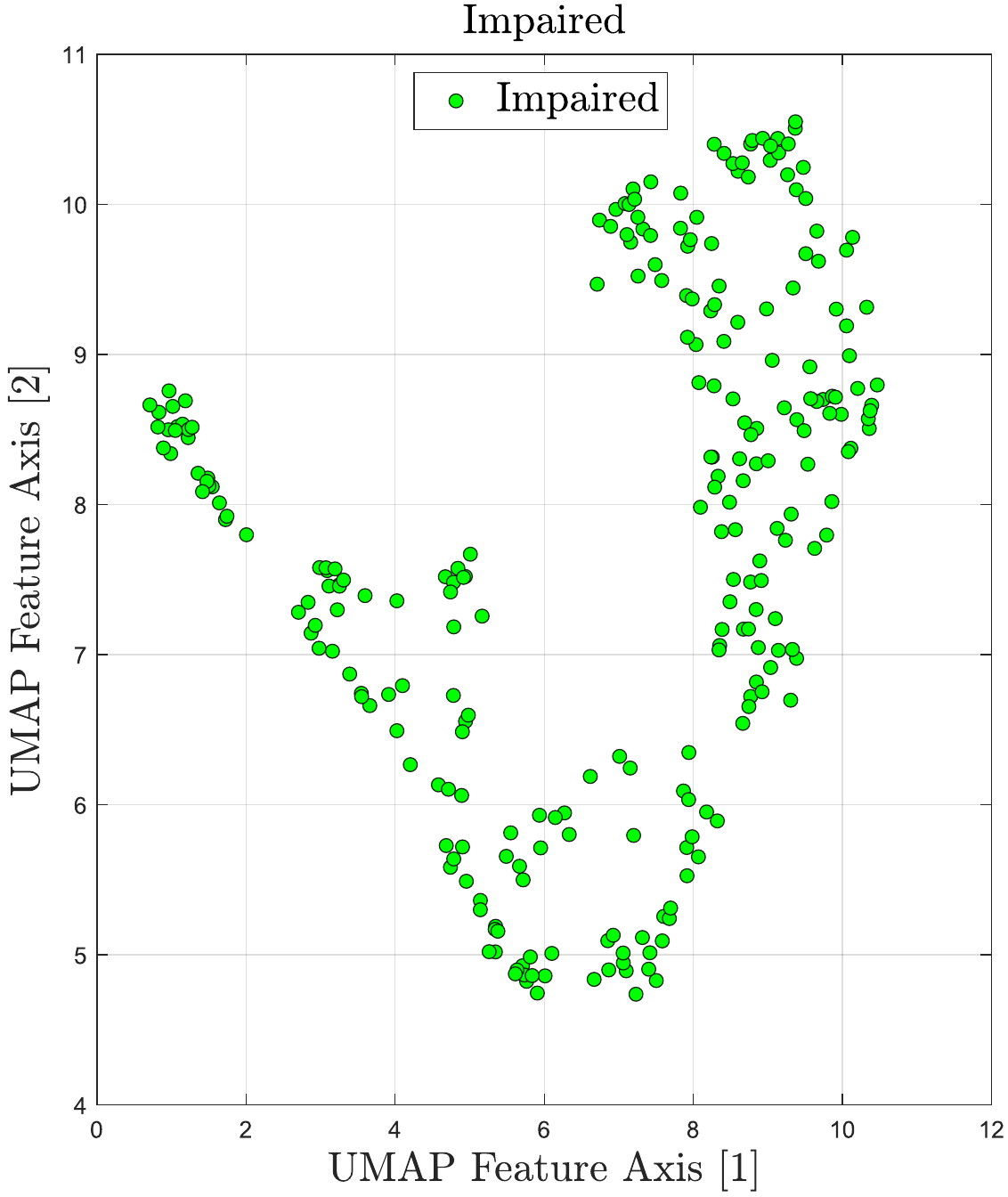}
         \caption{Impaired}
         \label{Impaired}
     \end{subfigure}
     \caption{UMAP feature space visualization}
     \label{fig:UMAP}
\end{figure}

\section{Results}
In this section, we presented the results of experiments conducted on the ImageNette, Cifar-10, and Cifar-100 datasets using ViTs, GANs, and CNNs under various fading and noise conditions, including Rayleigh, Rician, Nakagami-m fading, and AWGN. The ViT base model utilized in these tests was set up with a 16$\times$16 patch size, a 768 embedding dimension, and 12 transformer layers in both the encoder and decoder. The encoder used 12 self-attention heads, whereas the decoder had 16 heads. In addition, a DAE-ViT architecture was created, which includes a Rayleigh channel model with a noise factor of 0.2 to simulate channel degradation.
\begin{figure}[t!]
    \centering
    \includegraphics[width=\columnwidth]{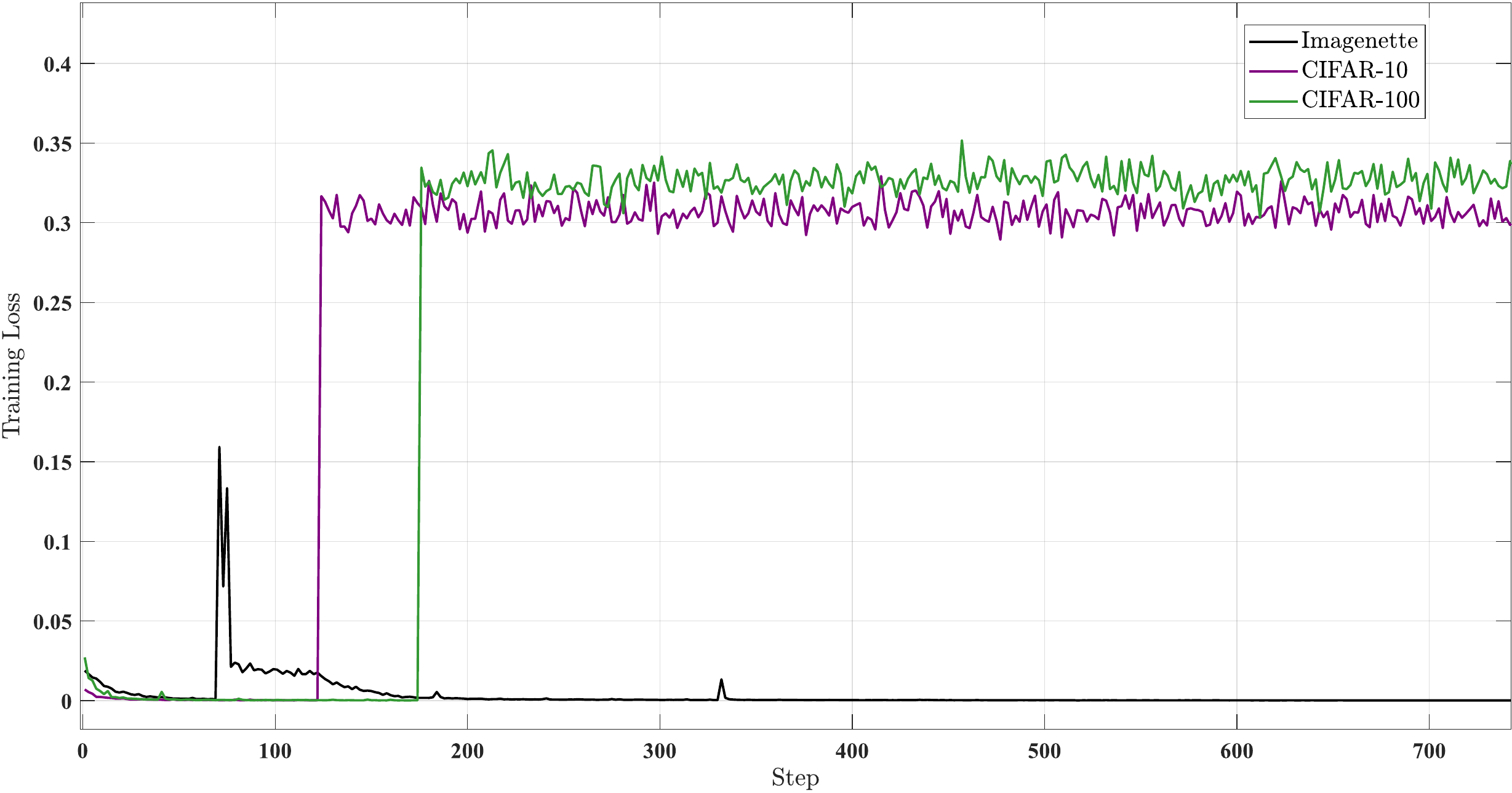}
    \caption{Train-loss curves on proposed data sets}
    \label{fig:train}
\end{figure}

Based on the training loss over multiple episodes, we evaluated the  Structural Similarity Index (SSIM) and Peak Signal-to-Noise Ratio (PSNR) Fig.~\ref{PSNR} and Fig.~\ref{SSIM} presents the PSNR and SSIM results respectively, where the ViT performed better than the CNN and GAN models across all datasets, reaching a PSNR of about 38 dB after 900 episodes. In contrast, the GAN and CNN models, had leveled off around 25 dB and 10 dB, respectively. The huge gap in PSNR values shows that the ViT outperformed the GAN and CNN models showing the ViT’s better ability for image reconstruction of semantic information. SSIM results also support the ViT’s higher efficiency. The ViT achieved an SSIM value near to 1.0 showing the proportion of structural similarity between the transmitted and recovered images was near-perfect, while the GAN and CNN models received lower SSIM values. 

Uniform Manifold Approximation and Projection (UMAP) feature space visualization indicates that the ViT clearly distinguishes between encoded and impaired features as shown in Fig.~\ref{Encoded} and Fig.~\ref{Impaired} respectively, demonstrating its capacity to effectively preserve semantic information even under demanding channel conditions.

\begin{table}[t!]
    \centering
    \caption{Performance Metrics for Various Fading and Noise Conditions}
    \label{tab:performance_metrics}
    \renewcommand{\arraystretch}{1.8}
    \resizebox{\columnwidth}{!}{%
        \begin{tabular}{|c|c|c|c|c|}
            \hline
            & \textbf{Rayleigh Fading} & \textbf{Rician Fading} & \textbf{Nakagami-\textit{m} Fading} & \textbf{Additive White Gaussian Noise} \\
            \hline
            \hline
            \textbf{Imagenette - Transformer} & 98.53 & 97.89 & 99.01 & 99.13 \\
            \textbf{Cifar-10 - Transformer} & 95.21 & 97.18 & 97.83 & 97.61 \\
            \textbf{Cifar-100 - Transformer} & 95.83 & 96.71 & 96.43 & 97.81 \\
            \textbf{Imagenette - GAN} & 95.21 & 94.22 & 94.71 & 94.83 \\
            \textbf{Cifar-10 - GAN} & 92.68 & 93.33 & 94.11 & 94.29 \\
            \textbf{Cifar-100 - GAN} & 92.01 & 92.56 & 92.97 & 93.17\\
            \textbf{Imagenette - CNN} & 92.56 & 93.87 & 93.54 & 93.79 \\
            \textbf{Cifar-10 - CNN} & 89.77 & 90.7 & 90.52 & 90.38 \\
            \textbf{Cifar-100 - CNN} & 89.24 & 90.23 & 90.16 & 90.01 \\
            \hline
        \end{tabular}%
    }
\end{table}
Table~\ref{tab:performance_metrics}
presents an overview of the performance metrics of the ViT model under various fading and noise conditions. The ViT model is more robust than the GAN and CNN models. Under Rayleigh fading, it achieved a PSNR of 98.53 dB on the ImageNette dataset whereas the GAN and CNN models obtained PSNR of 95.21 dB and 89.77 dB, respectively. Similar trends were seen in the Cifar-10 and Cifar-100 datasets.Fig.~\ref{fig:train} The loss curves for CIFAR-10 and CIFAR-100 exhibit more variability compared to Imagenette, suggesting dataset-specific differences that could influence training stability. 

\begin{figure}[t!]
    \centering
    \includegraphics[width=0.44\textwidth]{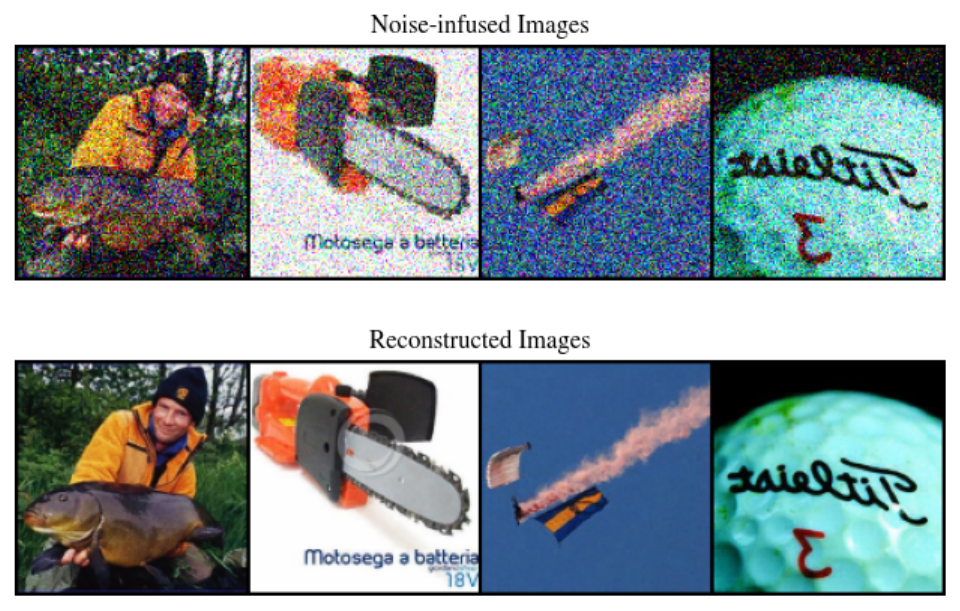}
    \caption{Output of reconstructed image}
    \label{fig:output}
\end{figure}

The ViT base model performed exceptionally well under Nakagami-m fading conditions, reaching PSNR values near or above 97 dB on all datasets, while the GAN and CNN models found it difficult to maintain comparable accuracy, especially on the Cifar-100 dataset. As it can be seen in Fig.~\ref{fig:output} the output was reconstructed with the exclusion of noise. Our results demonstrate that the proposed ViT architecture achieves a higher PSNR up to 38 dB, and SSIM values close to 1, while consuming 72\% less bandwidth

\section{Conclusion and Future Work}
This work presents the application of ViTs in semantic communication and exposes its contrast to traditional models like CNNs and  GANs. In this work, we propose a novel ViT architecture that delivers superior performance metrics, particularly in terms of PSNR and SSIM, with an emphasis on noise robustness and efficiency under varied noisy scenarios. Furthermore, proposed ViT architecture achieves a higher PSNR up to 38 dB, and SSIM values close to 1, while consuming 72\% less bandwidth. These findings highlight the significant potential of ViTs in communication systems.
In the future, we aim to explore hybrid models with noise reduction and optimization. While ViTs excel in semantic integrity, their resource demands limit use in under-resourced environments. Optimization techniques can enable edge device deployment.

\bibliographystyle{ieeetr}
\bibliography{references/main}

\end{document}